\documentclass[twocolumn,showpacs,preprintnumbers,amsmath,amssymb,APSl,prd,nofootinbib,superscriptaddress]{revtex4-1}

\usepackage{dcolumn}
\usepackage{bm}
\usepackage{ifpdf}
\usepackage{hyperref}
\usepackage{dcolumn}
\usepackage{bm}
\usepackage[spanish,english]{babel}
\usepackage{amsfonts}
\usepackage{amssymb}
\usepackage{graphicx}
\usepackage[latin1]{inputenc}

\newcommand{\LL}{\mathcal{L}}

\newcommand{\be}{\begin{equation}}
\newcommand{\en}{\end{equation}}
\newcommand{\bea}{\begin{eqnarray}}
\newcommand{\ena}{\end{eqnarray}}

\begin{document}

\title{Planck scale physics and topology change through an exactly solvable model}

\author{Francisco S. N.~Lobo}\email{flobo@cii.fc.ul.pt}
\affiliation{Centro de Astronomia e Astrof\'{\i}sica da
Universidade de Lisboa, Campo Grande, Ed. C8 1749-016 Lisboa,
Portugal}
\author{Jesus Martinez-Asencio}
\affiliation{Departamento de F\'{i}sica Aplicada, Facultad de Ciencias, Fase II, Universidad de Alicante, Alicante E-03690, Spain }
\email{jesusmartinez@ua.es}
\author{Gonzalo J. Olmo} \email{gonzalo.olmo@csic.es}
\affiliation{Departamento de F\'{i}sica Te\'{o}rica and IFIC, Centro Mixto Universidad de
Valencia - CSIC. Universidad de Valencia, Burjassot-46100, Valencia, Spain}
\author{D. Rubiera-Garcia} \email{drubiera@fisica.ufpb.br}
\affiliation{Departamento de F\'isica, Universidade Federal da
Para\'\i ba, 58051-900 Jo\~ao Pessoa, Para\'\i ba, Brazil}


\date{\today}

\begin{abstract}
We consider the collapse of a charged radiation fluid in a Planck-suppressed quadratic extension of General Relativity (GR) formulated \`{a} la Palatini. We obtain exact analytical solutions that extend the charged Vaidya-type solution of GR, which allows to explore in detail new physics at the Planck scale. Starting from Minkowski space, we find that the collapsing fluid generates wormholes supported by the electric field.  We discuss the relevance of our findings in relation to the quantum foam structure of space-time and the meaning of curvature divergences in this theory.
\end{abstract}

\maketitle

{\it Introduction.} An outstanding question in quantum gravitational physics is whether large metric fluctuations at the Planck scale may induce a change in topology. In fact, as suggested by Wheeler, at scales below the Planck length, the highly nonlinear and strongly interacting metric fluctuations may endow space-time with a foam-like structure \cite{Wheeler}. Thus, this hints that the geometry, and the topology, may be constantly fluctuating, so that space-time may take on all manners of nontrivial topological structures, such as wormholes \cite{Misner:1957mt}. However, one does encounter a certain amount of criticism to Wheeler's notion of space-time foam, for instance, in that stability considerations may place constraints on the nature or even on the existence of Planck-scale foam-like structures \cite{Redmount:1992mc}. Nevertheless, the notion of space-time foam is generally accepted, in that this picture leads to topology-changing quantum amplitudes and to interference effects between different space-time topologies \cite{Visser}, although these possibilities have met with some disagreement \cite{dewitt}.

Due to the multiply-connected nature of wormholes, their respective creation/generation inevitably involves the problematic issue of topology change \cite{acausal,Visser:1989ef}. The possibility that inflation might provide a natural mechanism for the enlargement of Planck-size wormholes to macroscopic size has been explored \cite{Roman:1992xj}. In fact, the construction of general relativistic traversable wormholes, with the idealization of impulsive phantom radiation, has been considered extensively in the literature \cite{Gergely02,Hayward02,Hayward04}.
Another problematic aspect in wormhole physics is that these geometries violate the pointwise energy conditions \cite{Morris}. However, this issue may be avoided in modified gravity, where the normal matter threading the wormholes may in principle be imposed to satisfy the energy conditions, and it is the higher order curvature terms that sustain these geometries \cite{Harko:2013yb}.
In fact, the general approach to wormhole physics is to run the gravitational field equations in the reverse direction, namely, consider first an interesting space-time metric and then through the field equations, the distribution of the stress-energy tensor components is deduced. However, one may rightly argue that this approach in solving the field equations lacks physical justification, and that a more physical motivation would be to consider a plausible distribution of matter-energy.

In this work we follow the latter route and find that in a quadratic extension of GR  formulated in the Palatini formalism wormholes can be generated dynamically. This result follows by probing Minkowski space-time with a charged null fluid, thus producing a Vaidya-type configuration. Now, switching off the flux, the metric settles down into a static configuration where the existence of a wormhole geometry becomes manifest. The topologically nontrivial character of the wormhole allows us to define the electric charge in terms of lines of electric force trapped in the topology. The size of the wormhole changes in such a way that the density of lines of force at the throat is given by a universal quantity. These facts allow to consistently interpret these solutions as geons in Wheeler's sense \cite{Wheeler} and has important consequences for the issue of the foam-like structure of space-time \cite{Misner:1957mt}.

{\it Dynamical charged fluid in Ricci-squared Palatini theories.}  Consider a theory defined by the action \cite{or12a}
\begin{equation}\label{eq:action}
S[g,\Gamma,\psi_m]=\frac{1}{2\kappa^2}\int d^4x \sqrt{-g}f(R,Q) +S_m[g,\psi_m]  \ ,
\end{equation}
where $\LL_G=f(R,Q)/(2\kappa^2)$ represents the gravity Lagrangian, $\kappa^2$ is a constant with suitable dimensions (in GR, $\kappa^2 \equiv 8\pi G/c^3$), $g_{\mu\nu}$ is the space-time metric,  $R=g^{\mu\nu}R_{\mu\nu}$, $Q=g^{\mu\alpha}g^{\nu\beta}R_{\mu\nu}R_{\alpha\beta}$, $R_{\mu\nu}={R^\rho}_{\mu\rho\nu}$,  ${R^\alpha}_{\beta\mu\nu}=\partial_{\mu}
\Gamma^{\alpha}_{\nu\beta}-\partial_{\nu}
\Gamma^{\alpha}_{\mu\beta}+\Gamma^{\alpha}_{\mu\lambda}\Gamma^{\lambda}_{\nu\beta}-\Gamma^{\alpha}_{\nu\lambda}\Gamma^{\lambda}_{\mu\beta}$
is the Riemann tensor of the connection $\Gamma^{\lambda}_{\mu\nu}$, and $S_m[g,\psi_m]$ represents the matter action (with $\psi_m$ the matter fields). We work in the Palatini formalism, where $g_{\mu\nu}$ and $\Gamma^{\lambda}_{\mu\nu}$ are regarded as independent fields. Setting the torsion to zero for simplicity, the field equations imply that $\Gamma^{\lambda}_{\mu\nu}$ can be written as the Levi-Civita connection of a metric $h_{\mu\nu}$ defined as \cite{or13a}
\begin{equation} \label{eq:h-g}
h^{\mu\nu}=\frac{g^{\mu\alpha}{\Sigma_{\alpha}}^\nu}{\sqrt{\det \hat{\Sigma}}} \ , \quad
h_{\mu\nu}=\left(\sqrt{\det \hat{\Sigma}}\right){{\Sigma^{-1}}_{\mu}}^{\alpha}g_{\alpha\nu} \ .
\end{equation}
We have defined the matrices $\hat\Sigma$ and $\hat P$, whose components are ${\Sigma_\alpha}^{\nu}\equiv \left(f_R \delta_{\alpha}^{\nu} +2f_Q {P_\alpha}^{\nu}\right)$ and ${P_\mu}^\nu\equiv R_{\mu\alpha}g^{\alpha\nu}$, with $f_X\equiv df/dX$. From the metric variation, one finds
\begin{equation}
2f_Q\hat{P}^2+f_R \hat{P}-\frac{f}{2}\hat{I} = \kappa^2 \hat{T} \label{eq:met-varRQ2} \ ,
\end{equation}
which represents a quadratic algebraic equation for ${P_\mu}^\nu$ as a function of ${[\hat{T}]_\mu}^\nu\equiv T_{\mu\alpha}g^{\alpha\nu}$. This implies that $R={[\hat{P}]_\mu}^\mu$, $Q={[\hat{P}^2]_\mu}^\mu$, and
${\Sigma_\alpha}^{\nu}$ are just functions of the matter sources. With elementary algebraic manipulations, Eq. (\ref{eq:met-varRQ2}) can be cast as
\begin{equation} \label{eq:fieldequations}
{R_{\mu}}^{\nu}(h)=\frac{\kappa^2}{\sqrt{\det \hat{\Sigma}}}\left(\LL_G\delta_{\mu}^{\nu}+  {T_\mu}^{\nu} \right) \ .
\end{equation}
This representation of the metric field equations implies that $h_{\mu\nu}$ satisfies a set of GR-like second-order field equations, with the right-hand side being determined by $\hat{T}$. Since $h_{\mu\nu}$ and $g_{\mu\nu}$ are algebraically related, it follows that $g_{\mu\nu}$ also verifies second-order equations. We note that whenever ${T_\mu}^{\nu}=0$, Eq. (\ref{eq:fieldequations}) recovers the vacuum Einstein equations \cite{or13a,lor13} (with possibly a cosmological constant, depending on the form of $\LL_G$) which, similarly as in other  Palatini theories \cite{Banados, Deser, Olmo:2013gqa}, guarantees the absence of ghost-like instabilities.

Let us consider the matter sector as described by a spherically symmetric flux of ingoing charged matter with a stress-energy tensor given by
\begin{equation}
T_{\mu\nu}^{flux}= \rho_{in} k_\mu k_\nu   \,,
\label{eq:Tmn-Max1}
\end{equation}
where $k_{\mu}$ is a null vector, satisfying $k_{\mu}k^{\mu}=0$, and $\rho_{in}$ is the energy density of the flux. The electric field generated by this flux contributes to the total stress-energy by means of
$T_{\mu\nu}^{em}=\frac{1}{4\pi}\left[F_{\mu\alpha}{F_{\nu}}^\alpha-\frac{1}{4}F_{\alpha\beta}F^{\alpha\beta} g_{\mu\nu}\right] \label{eq:Tmn-Max0} $.
If we consider a line element of the form
\begin{equation}\label{eq:ds2g}
ds^2=-A(x,v) e^{2\psi(x,v)}dv^2+ 2e^{\psi(x,v)}dv dx+r^2(v,x)d\Omega^2 \ ,
\end{equation}
then Maxwell's equations, $\nabla_\mu F^{\mu\nu}=4\pi J^\nu$, where $J^\nu\equiv \Omega(v) k^\nu$ is the current of the ingoing flux, lead to
 $r^2 e^{\psi(x,v)}F^{xv}=q(v)$, where $q(v)$ is an integration function. These equations also imply that $\Omega(v)\equiv q_v/4\pi r^2$.

To proceed further, we focus on a simple quadratic extension of GR,
\begin{equation}\label{eq:quadratic}
f(R,Q)=R+l_P^2(a R^2+Q) \ ,
\end{equation}
where $l_P \equiv \hbar G/c^3$ is the Planck length and $a$ is a free parameter.
Having specified the matter sources and the gravity Lagrangian, one finds $R=0$, $Q=\kappa^2 q^4/4\pi r^8$, and
\begin{equation} \label{eq:Sigma}
{\Sigma_{\mu}}^{\nu}=\left(
\begin{array}{cccc}
\sigma_-  & \sigma_{in} & 0 & 0 \\
0 & \sigma_- & 0 & 0 \\
0 & 0 & \sigma_+ & 0 \\
0 & 0 & 0 & \sigma_+  \\
\end{array}
\right) ,
\end{equation}
where $\sigma_{\pm}=1\pm \frac{{\kappa}^2 l_P^2 q^2(v)}{4\pi r^4}$ and $\sigma_{in}=\frac{2\kappa^2 l_P^2 \rho_{in}}{1-2{\kappa}^2 l_P^2 q^2(v)/4\pi r^4}$. The field equations (\ref{eq:fieldequations}) can thus be written as
\begin{equation}\label{eq:Rmn}
{R_{\mu}}^{\nu}(h)=\left(
\begin{array}{cccc}
-\frac{{\kappa}^2q^2(v)}{8\pi r^4\sigma_+} & \frac{e^{-\psi}\kappa^2\rho_{in}}{\sigma_+\sigma_-}  & 0 & 0  \\
0 & -\frac{{\kappa}^2q^2(v)}{8\pi r^4\sigma_+} & 0 & 0  \\
0& 0& \frac{{\kappa}^2q^2(v)}{8\pi r^4\sigma_-} & 0 \\
0& 0& 0 & \frac{{\kappa}^2q^2(v)}{8\pi r^4\sigma_-}
\end{array}
\right)  \ .
\end{equation}
 The strategy now is to solve for $h_{\mu\nu}$ first and then use (\ref{eq:h-g}) and (\ref{eq:Sigma}) to obtain $g_{\mu\nu}$. For this purpose, we define a line element for $h_{\mu\nu}$ of the form $d\tilde{s}^2= -F(v,x)e^{2\xi(v,x)}dv^2+2e^{\xi(v,x)}dv dx + \tilde{r}^2(v,x)d\Omega^2$. The field equations then imply that we can take $\tilde{r}(v,x)=x$, $e^{\xi(v,x)}=1$, and $F(x,v)=1-2M(x,v)/x$, with $M(x,v)$ satisfying
\begin{equation}\label{eq:Mx-Mv}
M_x=\frac{{\kappa}^2q^2}{16\pi  r^2} \  , \ M_v=\frac{\kappa^2\rho_{in} r^2}{2} \ ,
\end{equation}
 where the relation (\ref{eq:h-g}) establishes that
\begin{equation}
r^2(x,v)=\frac{x^2+\sqrt{x^4+l_P^2 {\kappa}^2q^2(v)/\pi}}{2} \ . \label{eq:rtx0}
\end{equation}
By direct integration of $M_x$, we find
\begin{equation}\label{eq:mass}
M(x,v)=M_0+\gamma(v)+\frac{\tilde{\kappa}^2q^2(v)}{4}\int \frac{dx}{r^2}\Big|_{v=\text{const}} \ ,
\end{equation}
where $M_0$ is a constant and $\gamma(v)$ an integration function.
Computing $M_v$ and comparing with Eq. (\ref{eq:Mx-Mv}), one finds
\begin{equation}
\rho_{in} r^2=\frac{2}{\kappa^2} \left[\gamma_v + \frac{\kappa^2 q q_v}{8\pi}\int \frac{dx}{r^2\sigma_+} \right] \ , \label{eq:constraint}
\end{equation}
which is fully consistent with $\nabla_\mu (T^{\mu\nu}_{flux}+T^{\mu\nu}_{em})=0$.

We have thus found a complete solution to the Palatini theory  (\ref{eq:quadratic}) coupled to a stream of charged null radiation. In this solution, $q_v\equiv \partial_v q(v)$ and $\gamma_v\equiv \partial_v \gamma(v)$ are free functions. Given the structure of the mass function in Eq. (\ref{eq:mass}) and to make contact with previous results on static configurations \cite{or12a} of the same model (\ref{eq:quadratic}) , we find it useful to write it as
\begin{equation}
M(x,v)= M_0+\gamma(v) +\frac{r_q(v)^2}{4r_c(v)}\left(\int dz G_z\right)\Big|_{z=r/r_c} \ ,
\end{equation}
where $r_q^2(v)\equiv \kappa^2 q^2/4\pi$, $r_c(v)\equiv\sqrt{r_q(v)l_P}$, $z(x,v)\equiv r(x,v)/r_c(v)$ and $G_z=\frac{z^4+1}{z^4 \sqrt{z^4-1}}$, where we have used the relation $dr/dx=\sigma_{-}^{1/2}/\sigma_{+}$ (at constant $v$). The function $G(z)$ can be written as an infinite power series and its form was given in \cite{or12a}. Using the compact notation
\begin{equation}
M(x,v)=M(v)\left[1+\delta_1(v) G\left(z\right)\right] \ ,
\end{equation}
where $M(v)=M_0+\gamma(v)\equiv r_S(v)/2 $ and $\delta_1(v)=\frac{1}{2r_S(v)} \sqrt{\frac{r_q(v)^3}{l_P}}$, the line element (\ref{eq:ds2g}) becomes
\begin{eqnarray}
ds^2&=&-\left[\frac{1}{\sigma_+}\left(1-\frac{1+\delta_1 (v) G(z)}{\delta_2(v) z \sigma_{-}^{1/2}}\right)- \frac{2l_P^2 \kappa^2 \rho_{in}}{\sigma_{-}(1-\frac{2r_c^4}{r^4})}\right]dv^2
    \nonumber\\
 &&+ \frac{2}{\sigma_+}dvdx+r^2(x,v) d\Omega^2 \ , \label{eq:ds2final}
\end{eqnarray}
where $\delta_2(v)\equiv \frac{r_c(v)}{r_S(v)}$. Equation (\ref{eq:ds2final}) constitutes the main result of this paper.

\begin{figure}[h]
\includegraphics[width=0.45\textwidth]{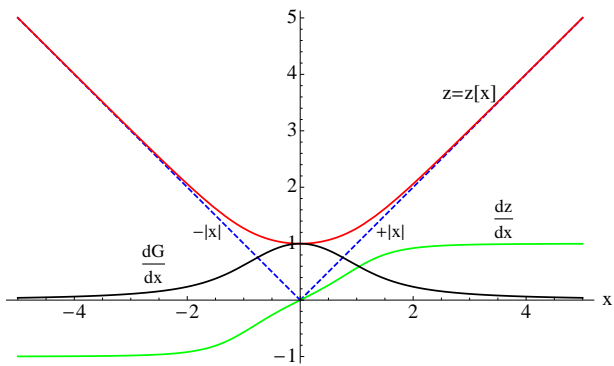}
\caption{The minimum of $z(x)$ implies the existence of a wormhole extension of the geometry, with $x$ covering the whole real axis  $-\infty <x<+\infty$. Note the smoothness of $dG/dx$ at $x=0$. In this plot, $r_c=1$.
\label{fig:WH_extension}}
\end{figure}

{\it Dynamic generation of wormholes.} Consider now that the initial state is flat Minkowski space and assume that a charged perturbation of compact support propagates within the interval $[v_i,v_f]$.  Given the relation (\ref{eq:rtx0}), which can be written as
$r^2(x,v)=\frac{x^2+\sqrt{x^4+4r_c^4(v)}}{2}$, 
it follows that for $v<v_i$ we have $r^2(x,v)=x^2$, which extends from zero to infinity. Entering the $v\ge v_i$ region, this radial function, which measures the area of the $2$-spheres of constant $x$ and $v$, never becomes smaller than $r_c^2 (v)$, where this minimum is reached at $x=0$. If we now consider the region $v>v_f$, in which $\rho_{in}$ is again zero, the result is a static geometry.
One can verify \cite{or12a} that in this static geometry curvature scalars generically diverge at $x=0$ except if the charge-to-mass ratio $\delta_1^f \equiv \delta_1(v\ge v_f)$ takes the value  $\delta_1^f=\delta_1^* \simeq 0.572$. This $\delta_1^*$  is a constant that appears in the series expansion of $G(z)=-1/\delta_1^*+2\sqrt{z-1}+\ldots$ as $z\to 1$. The smoothness of the geometry when $\delta_1^f=\delta_1^*$, together with the fact that $r(x)$ reaches a minimum at $x=0$, allow to naturally extend the coordinate $x$ to the negative real axis, thus showing that the radial function $r^2(x)$ bounces off to infinity as $x\to -\infty$ (see Fig.\ref{fig:WH_extension}).
This puts forward the existence of a wormhole structure with its throat located at $x=0$.

It is important to note that  the flux $\Phi= \int_S *F$, where $*F$ is the 2-form dual to the Faraday tensor, through any closed 2-surface $S$ enclosing  $x=0$
is non-zero and can be used to define the charge inside $S$, $\Phi=4\pi q(v)$, on a purely topological basis. One thus finds that the density of lines of force at $x=0$ is given by $\Phi/4\pi r_c^2(v)=\sqrt{c^7/2\hbar G^2}$, which is constant and  independent of the particular values of the charge and mass of the solution. This shows that the (topological) wormhole structure exists even when $\delta_1(v)\neq \delta_1^*$ and, therefore, is insensitive to the existence of local curvature divergences.

We also note that for $x\gg r_c(v)$, the line element (\ref{eq:ds2final}) quickly recovers the expected GR prediction in both the dynamic and the static case. Event horizons are thus expected in general, though there exist nonsingular naked configurations with $\delta_1^f=\delta_1^*$ \cite{lor13,or12a}, which represent genuine  wormholes.

{\it Discussion and Summary.}
The dynamical generation of wormholes outlined above, in the context of charged fluids in quadratic Palatini gravity, differs radically in nature relative to the construction of general relativistic traversable wormholes, with the idealization of impulsive phantom radiation considered in the literature \cite{Gergely02,Hayward02,Hayward04}. More specifically, in the latter, it was shown that the adequate synchronization of the energies and the emission timing of two opposing streams of phantom radiation may support a static traversable wormhole \cite{Hayward02}. The theory presented here generates static wormholes by means of a finite pulse of electrically charged radiation, without the need to keep two exotic-energy streams active continuously or to synchronize them across the wormhole.

Regarding the size of the wormholes, we note that if instead of using $l_P^2=\hbar G/c^3$ to characterize the curvature corrections one considers a different length scale, say $l_\epsilon^2$, then their area would be given by $A_{WH}=\left(\frac{l_\epsilon}{l_P}\right)\frac{2N_q}{N_q^c}A_P$, where $A_P=4\pi l_P^2$, $N_q=|q/e|$ is the number of charges, and $N_q^c\approx 16.55$. Though this could allow to reach sizes orders of magnitude larger than the Planck scale, it does not seem very likely that macroscopic wormholes could arise from any viable theory of this form, though the effects of other matter/energy sources might be nontrivial.

The meaning and implications of classical singularities has been a subject of intense debate in the literature for years. Their existence in GR is generally interpreted as a signal of the limits of the theory, where the quantum effects of gravity should become relevant and an improved theory would be necessary. This is, in fact, the reason that motivates our  study of quadratic corrections beyond GR. As pointed out above and shown in detail in \cite{or12a}, the curvature divergences for the static wormhole solutions arising in quadratic Palatini gravity with electrovacuum fields are much weaker than their counterparts in GR (from $\sim 1/r^8$ in GR to $\sim (\delta_1-\delta_1^*)^2/(r-r_c)^3$ in our model). We have seen here that the existence of these curvature divergences is not an obstacle to have a well defined electric flux across the $r=r_c(v)$ surface. As a result,  the function $r^2$ never drops below the scale $r_c^2$ and  implies that the total energy stored in the electric field is finite (see \cite{Olmo:2013gqa,lor13} for details), which clearly contrasts with the infinite result that GR yields. Therefore, even though curvature scalars may diverge, physical magnitudes such as total mass, energy, and electric charge are insensitive to those divergences, which demands for a more in depth analysis of their meaning and implications.

In the dynamical case discussed here,  transient curvature divergences may also arise due to the $\rho_{in}$ term in the $g_{vv}$ component of the metric. A glance at (\ref{eq:ds2final}) indicates that there might be divergences at $r=r_c(v)$, where $\sigma_-$ vanishes, and at $r^4=2r_c^4(v)$. However, if one considers the case $\delta_1(v)=\delta_1^*$, which avoids the divergences in the static case, a short algebra confirms that the geometry is smooth at $r=r_c(v)$ also in the dynamical case, with $\rho_{in}\to 0$ there. Therefore, no large metric fluctuations occur at the wormhole throat  for this choice of $\delta_1(v)$. At $r=2^{1/4}r_c(v)$, a divergence occurs, though it disappears when the infalling flux ceases. The details of this transient will be discussed elsewhere \cite{inprogress}.

We note that since in our theory the field equations outside the matter sources recover those of vacuum GR, Birkhoff's theorem must hold in those regions. This means that for $v<v_i$ we have Minkowski space, whereas for $v>v_f$ we have a Reissner-Nordstr\"om-like geometry. The departure from Reissner-Nordstr\"om is due to the Planck scale corrections of the Lagrangian, which are excited by the presence of an electric field, and only affect the microscopic structure, which is of order $\sim r_c(v)$.  Due to the spherical symmetry and the second-order character of the field equations, Birkhoff's theorem guarantees the staticity of the solutions for $v>v_f$.

In concluding, we have found an exact analytical solution for the dynamical process of collapse of a null fluid carrying energy and electric charge in a quadratic extension of GR formulated \`{a} la Palatini. This extends the well-known Vaidya-Bonnor solution of GR \cite{BV} to a new scenario that allows to explore in detail new physics at the Planck scale. Focusing on the initial and final static configurations, we have shown that wormholes can be formed out of Minkowski space by means of a pulse of charged radiation, which contrasts with previous approaches in the literature requiring artificial configurations and synchronization of two opposing streams of phantom energy.

Though we have considered an idealized pressureless and ultra-relativistic charged fluid in a highly symmetric scenario,  we believe that our results support the view that space-time could have a foam-like microstructure with wormholes generated by  fluctuations of the quantum vacuum involving the spontaneous creation/anihiliation of very energetic charged particles. In fact, all spherically symmetric charged solutions of our theory able to extend their electric field down to scales  of order $r_c\sim l_P$ possess a wormhole structure. We understand that  pressure and other dispersion effects should act so as to prevent the effective concentration of charge and energy on scales of order  $\sim l_P$, thus suppressing wormhole production in low energy scenarios.

Let us finally stress that, in general, the wormhole structures found here develop curvature divergences but even so are characterized by well-defined and finite electric charge and total energy \cite{Olmo:2013gqa,lor13}. The physical role that such divergences could play is thus uncertain and requires an in-depth analysis. To fully understand these issues, our model should be improved to include several important aspects, such as the presence of gauge fields and/or fermionic degrees of freedom, or to consider the dynamics of counter-streaming effects due to the presence of simultaneous ingoing and outgoing fluxes. Several lines of research in this and other directions are currently underway.

{\it Acknowledgments.}
FSNL acknowledges financial support of the Funda\c{c}\~{a}o para a Ci\^{e}ncia e Tecnologia through an Investigador FCT Research contract, with reference IF/00859/2012, funded by FCT/MCTES (Portugal), and grants CERN/FP/123615/2011 and CERN/FP/123618/2011. GJO is supported by the Spanish grant FIS2011-29813-C02-02, the Consolider Programme CPAN (CSD2007-00042), and the JAE-doc program of the Spanish Research Council (CSIC). DRG is supported by CNPq (Brazilian agency) through project No. 561069/2010-7 and thanks the hospitality and partial support of the Theoretical Physics Department of the University of Valencia.

\end{document}